\documentclass[english, 12pt]{article}
\usepackage[T1]{fontenc}
\usepackage[latin9]{inputenc}
\usepackage[a4paper,margin=2cm,top=2cm,bottom=2cm]{geometry}
\usepackage{color}
\usepackage{amsmath}
\usepackage{graphicx}
\usepackage{amssymb}
\usepackage{esint}
\usepackage{babel}
\usepackage{float}
\newcommand{\lsim}{\mathrel{\hbox{\rlap{\lower.55ex \hbox{$\sim$}} \kern-.3em \raise.4ex \hbox{$<$}}}}
\newcommand{\aff}[1]{\par{\raggedright#1\vspace{1.4em}\noindent\par}}

\begin{document}

\title{Dark antiatoms can explain DAMA}
\date{}
\author{Quentin Wallemacq\thanks{quentin.wallemacq@ulg.ac.be}\ ~ and Jean-Ren\'e Cudell\thanks{jr.cudell@ulg.ac.be}}
\maketitle

\aff{\begin{center} IFPA, AGO Department, University of Li\`ege, Sart Tilman, 4000 Li\`ege, Belgium \par 
\end{center}}

\begin{abstract}

We show that the existence of a sub-dominant form of dark matter, made of dark   ``antiatoms'' of mass $m\sim 1$ 
~TeV and size $\dot{a}_0\sim 30$~fm, can explain the results of direct detection experiments, with a positive 
signal in DAMA/NaI and DAMA/LIBRA 
and no signal in other experiments. The signal comes from the binding of the dark antiatoms to
 thallium, a dopant in DAMA, and is not present for the constituent atoms of other experiments. The dark
 antiatoms are made of two particles oppositely charged under a dark $U(1)$ symmetry and can bind to terrestrial 
 atoms because of a kinetic mixing between the   photon and the massless dark photon, such that the dark 
 particles acquire an electric millicharge $\sim \pm 5.10^{-4}e$. This millicharge enables them to bind to  high-$Z$ 
 atoms via radiative capture, after they thermalize in terrestrial matter through elastic collisions.

\end{abstract}

\section{Introduction}
Direct dark matter experiments such as XENON100 \cite{Aprile:2012nq}, LUX \cite{Akerib:2013tjd}, CDMS-II/Ge 
\cite{Ahmed:2010wy} and superCDMS \cite{Agnese:2014aze} have reported the non-observation of nuclear recoils, 
and the subsequent bounds on the cross section of dark matter colliding with nuclei. A few experiments have reported 
positive signals, but the probability for the signal to be due to background is 5\% in the case of CDMS-II/Si \cite{Agnese:2013rvf}, a 
possible misinterpretation of  surface events at near-threshold energies \cite{Davis:2014bla} may weaken the results 
of CoGeNT \cite{Aalseth:2014eft} , and the CRESST II collaboration does not confirm the excess it previously saw 
\cite{Angloher:2014myn}. It thus seems reasonable to assume  that
DAMA is the only positive direct search experiment. The combined DAMA/NaI and DAMA/LIBRA signal has 
a significance of $9.3~\sigma$ level \cite{Bernabei:2013xsa} and although the origin of the signal is not certain, its
existence is beyond any doubt. 

Producing a signal in DAMA and only in DAMA is not easy. Clearly, Weakly Interacting Massive Particles (WIMPs) 
scattering off nuclei within the detectors will not discriminate between detectors (see Ref. \cite{Panci:2014gga} for a 
review). A number of alternative models  have been proposed, with different interaction mechanisms 
\cite{Panci:2014gga,Foot:2012rk,Cline:2012is,Khlopov:2010ik,Cudell:2012fw,McCullough:2013jma,Wallemacq:2013hsa,Wallemacq:2014lba} , 
that can produce a signal in 
some experiments and not in others. But so far no model restricts the signal only to DAMA.

Here we shall use a very simple model, based on the mixing of a dark sector with visible matter via a kinetic mixing 
term of a dark $U(1)$ with   photons. The net effect of such a mixing is that dark matter acquires a millicharge 
\cite{Holdom:1985ag} $\dot q=\epsilon e$ with $\epsilon \sim 0.5~10^{-3}$ (we shall use a dot to denote all the quantities 
related to dark matter).  We shall also assume that dark matter is made of hydrogen-like atoms, bound by the dark 
electromagnetic field. The signal in DAMA will come from the radiative capture of the dark nucleus, which thus must carry
a negative millicharge. We shall refer to it as a ``dark antiproton'', bound to a ``dark positron'' into a ``dark anti-hydrogen''. 
As it will turn out, the masses of dark species must roughly be 1000 times larger than that of their counterparts 
in visible matter. 

To avoid nuclear recoils, we need the dark antiatoms to thermalize in the Earth via elastic  scattering on   atoms, similarly to 
what was done in \cite{Wallemacq:2013hsa,Wallemacq:2014lba}. The dark antiatoms are almost at rest when they 
enter the detector, and they do not produce detectable nuclear recoils. Their non-negligible cross sections with visible
matter imply that they will have large self-interactions. They must thus be considered as a sub-dominant species
accounting for a fraction $f$ of the local dark matter density, in order to avoid the bounds  on self-interacting dark matter from halo shapes 
\cite{2002ApJ...564...60M}.

The DAMA signal would then be due to the radiative capture of the dark antiprotons  by conventional matter. It is well known that 
atomic capture cross sections are large and depend on the size of the atoms. To obtain capture cross sections that fall within the 
range required by DAMA, dark anti-hydrogen atoms of rather small size are needed,
 i.e. with a Bohr radius $\dot{a}_0\sim30$ fm. This signal
would  be counted as electronic recoil in other detectors. However, one needs to be careful, as a simple
translation of the DAMA modulation to LUX \cite{Akerib:2013tjd} or XENON 100 \cite{Aprile:2012nq} would exceed 
the total electronic background. Hence an extra ingredient is needed to make the DAMA detector special.

Because of the screening due to the charge distribution around the dark nucleus, the attractive potential well between a nucleus and a dark atom 
has a reduced size of the order of $\dot{a}_0$. It is then not guaranteed that opposite charges will have bound states. 
For the bound states to 
exist, the potential must be deep enough, i.e. the atomic nucleus needs to have a high enough $Z$.
Although DAMA is mostly made of sodium and iodine, it also contains traces of thallium (Z=81). This is much
heavier than the elements constituting the other detectors, so that it could be that dark antiatoms bind
to thallium, and that only DAMA observes a signal.

In Section \ref{darkSect}, we present the main ingredients of the model, consider the current constraints on millicharges and discuss the galactic 
distribution of the dark atoms. In Section \ref{Binding}, we derive the atom-dark atom interaction potential and show that binding is possible only 
with heavy elements. In Section \ref{secDAMArate}, we set the scenario from space to underground detectors by discussing the thermalization in 
the terrestrial crust and deriving the capture cross section as well as the event rate in an underground detector. The parameter space of the 
model is explored and models reproducing the results of DAMA at the $2\sigma$ level in full consistency with the negative results of the 
other experiments are given. We discuss the possible absorption by lead in the crust and in the shielding of the detector in Section \ref{AnElem}, 
and consider the limits on the abundance of anomalously heavy isotopes of known elements. We conclude by proposing simple ways to test the 
present model.


\section{Dark sector}\label{darkSect}

We take a very simple dark sector, in which two fermions $\dot{p}^{-}$ and $\dot{e}^{+}$, of masses $m_{\dot{p}^{-}}$ and 
$m_{\dot{e}^{+}}$ and of respective charges $-\dot{e}$ and $+\dot{e}$ under a dark $U(1)$, bind to each other and form dark anti-
hydrogen atoms. In that system, $\dot{p}^{-}$ plays the role of the nucleus, {\it i.e.} we assume $m_{\dot{p}^{-}}\gg m_{\dot{e}^{+}}$. 
The mass $m$ of the atom is $m\simeq m_{\dot{p}^{-}}$, its Bohr radius $\dot{a}_0$ is given by $\dot{a}_0=1/\left(\dot{\mu}
\dot{\alpha}\right)$, where $\dot{\mu}=m_{\dot{p}^{-}}m_{\dot{e}^{+}}/\left(m_{\dot{p}^{-}}+m_{\dot{e}^{+}}\right)\simeq 
m_{\dot{e}^{+}}$ is the reduced mass and $\dot{\alpha}=\dot{e}^{2}/4\pi$. The dark antiatoms form a self-interacting component of
Dark Matter. 

In order to produce a non-gravitational interaction between the dark and the visible sector, dark massless photons $\dot{\gamma}$ 
associated with the dark $U(1)$ are kinetically mixed with the   photon, similarly to \cite{Foot:2012rk,Cline:2012is,
Holdom:1985ag,Feldman:2007wj}, making $\dot{p}^{-}$ and $\dot{e}^{+}$ behave like electric millicharges of values $-\epsilon e$ and
$+\epsilon e$, where $\epsilon$ is the dimensionless mixing parameter. We assume that $\dot{\gamma}$ is massless to avoid several 
constraints holding for massive para-photons, {\it e.g.} from the anomalous magnetic dipole moments of the electron and the muon 
\cite{Cline:2012is}.

Note that for massless dark photons there are several equivalent definitions for the fields $\gamma$ and $\dot{\gamma}$.  One possibility is to 
keep the interaction between both sectors as the exchange of a photon that converts into a dark photon or conversely, each of them being 
coupled only to its own sector. 
Another possibility is to diagonalise the Hamiltonian by defining a  photon that couples to the visible current with $e$ and to the dark 
current with $\epsilon e$, while the dark photon couples only to the dark sector with $\dot{e}$. This means that particles charged under the 
dark $U(1)$ appear as millicharges in the visible sector. 
Ref. \cite{Essig:2013lka} summarizes the constraints on millicharged particles from direct laboratory tests and from cosmological and 
astrophysical observations. Very strong limits on $\epsilon$ can be obtained from the cooling of red giants and white dwarfs, from the alteration 
of the baryon-to-photon ratio during Big Bang Nucleosynthesis or from the invisible decay mode of orthopositronium into a pair of millicharged 
particles, but they hold all together for masses smaller than $1$ MeV, while we will be interested here by $m_{\dot{e}^{+}}\approx 1$ GeV and 
$m_{\dot{p}^{-}}\approx 1$ TeV. For that range of masses, direct bounds from accelerators leave a large allowed window with $\epsilon<0.1$ 
for masses $>1$ GeV.

An additional constraint comes from the disruptions of the acoustic peaks of the Cosmic Microwave Background (CMB) in the presence of 
millicharged particles \cite{Dolgov:2013una}. In view of the Planck data, it sets an upper limit on the cosmological density of millicharges 
$\Omega_{mc}h^2<0.001\,(95\%)$, but is assumes that the millicharged dark matter is fully ionized. This should be weakened here as the 
oppositely charged particles form neutral atomic structures while only an ionized fraction remains. Moreover, constraints on self-interacting dark 
matter from halo shapes \cite{2002ApJ...564...60M} and colliding clusters can be completely avoided \cite{Fan:2013yva}, if no more than 
$10\%$ of the whole dark matter in halos has self-interactions while the rest is collisionless. This is an additional weakening factor of the CMB 
constraint that leaves us with a subdominant sector to account for the results of the direct searches.

According to Ref. \cite{Fan:2013yva}, such a self-interacting dark matter with a cooling mechanism is likely to form dark disks in galaxies. The 
emission of dark photons by the atoms can contribute to this process and we expect dark antiatoms, similarly to baryons, to concentrate 
in a disk, aligned or not with that of visible matter. In that particular case, stellar velocities in and out of the galactic plane give stronger 
bounds on the amount of self-interacting dark
 matter and the observation of the kinematics of nearby stars leads to a limit on the mass of dark atoms of $5\%$ of the total dark mass of the 
 Milky Way halo. We will use the value $0.3$ GeV/cm$^{3}$ for the local dark matter density and consider that dark antiatoms 
 make a fraction $f
 \in]0,1]$ of it. The rest can be made of Cold Dark Matter (CDM) that only interacts gravitationally with visible matter and  does not produce 
 any signal in underground detectors. Note that the case $f=1$ is possible in case of quasi-alignment of the dark and baryonic disks, as a thin disk 
 can be much more concentrated than a diffuse halo.

While the velocity distribution of the dark particles in the halo cannot be neglected when the direct searches are interpreted in terms of collisions 
between nuclei and weakly interacting particles, it is much less important here since all the dark atoms thermalize in terrestrial matter and end up 
with the same thermal distribution. For that reason, we assume that the dark atoms are at rest in the frame of the dark disk, which itself is at 
rest with respect to the halo of collisionless particles.

\section{Binding to very heavy elements}\label{Binding}
The dark antiatoms, after they have lost most of their energy in terrestrial matter, bind to the atoms of the active medium of a detector. At 
long distances, the atoms and dark antiatoms are neutral, and the potential is zero. As the antiatom approaches, and neglecting van der Waals 
forces which are extremely small, the electron cloud  and the dark positron cloud start
to overlap, without any Fermi repulsion as the electrons and dark positrons are different particles. 
This causes first a rather weak repulsive force, but as the two systems get closer, an attractive 
 interaction of the nucleus with the negatively charged $\dot{p}^{-}$ develops at  distances close to the radius of the
dark antiatom, which will turn out be of the order of $30$ fm to reproduce the event rate of DAMA/LIBRA. The attractive force reaches a 
maximum at distances of the order of the radius of the nucleus. Hence we have a rather narrow attractive potential, shown in Fig. 1, 
which will be the source of the capture cross section.

The first condition is that bound states must exist, and it is well known that the narrower the potential, the deeper it has to be. So in this case 
not only is the effective charge reduced by $\epsilon$, but also the narrowness of the potential implies that it must be deep, {\it i.e.} that $Z$
must be large. The second condition will be that we capture the antiatom on a $p$ bound state, starting from an $s$ state of the continuum, so that it can emit a photon during an electric dipole transition. This in turn will require
larger values of $Z$.

To determine precisely the  interaction potential between the atom and the dark antiatom, we consider the four interacting 
charges. The millicharges are easy to model: we take the dark antiproton as a point charge $-\epsilon e$, and assume that the dark positron, of 
charge $\epsilon e$, is in a 1s hydrogen-like orbital. For the visible atom, we take its nucleus as a uniform charge distribution of radius
$R$(fm)$=1.2A^{1/3}$ and charge $+Ze$, and use explicit atomic form factors \cite{Formfact:2006} to treat the electron cloud.
As the interaction between atom and dark antiatom is rather weak, we assume that the charge structure is not modified during the
interaction. The total atom-dark atom interaction potential is then the sum of four terms: the nucleus-$\dot{p}^{-}$ potential 
$V_{N\dot{p}^{-}}$, the nucleus-$\dot{e}^{+}$ potential $V_{N\dot{e}^{+}}$, the electron-$\dot{p}^{-}$ potential $V_{e\dot{p}^{-}}$ and 
the electron-$\dot{e}^{+}$ potential $V_{e\dot{e}^{+}}$:
\begin{equation}
 V(r)=V_{N\dot{p}^{-}}(r)+V_{N\dot{e}^{+}}(r)+V_{e\dot{p}^{-}}(r)+V_{e\dot{e}^{+}}(r),\label{eq:1}
 \end{equation}
where $r$ is the distance between the center of the   nucleus and $\dot{p}^{-}$.

The first term in \eqref{eq:1} corresponds to the potential between a point charge and a uniform sphere, and is given by:
\begin{equation}
V_{N\dot{p}^{-}}(r)=-\frac{Z\epsilon\alpha}{2R}\left(3-\frac{r^{2}}{R^{2}}\right)\theta(R-r)-\frac{Z\epsilon\alpha}{r}\theta(r-R)\label{eq:2}
\end{equation}
The other terms involve diffuse distributions and can be calculated through the use of  form factors. It is shown in Appendix \ref{appendixA} that 
the Fourier transform $\tilde V(\vec q)$ of the electrostatic potential $V(\vec D)$ between two charge distributions $\rho_1(\vec x)$ and $
\rho_2(\vec y)$ is related to their Fourier transforms $F_1(\vec q)$ and $F_2(\vec q)$ , i.e. to their form factors, through:
\begin{equation}
\tilde V(\vec q)=\frac{F_1(\vec q)F_2(-\vec q)}{q^2}\label{eq:3},
\end{equation}
where $\vec D$ is the position vector between two arbitrary points taken in each distribution and where $\vec x$ and $\vec y$ locate 
respectively the charges in the distributions $1$ and $2$ with respect to these arbitrary points. The Fourier variable $\vec q$ is the transferred 
momentum, i.e. the momentum of the exchanged photon. When the distributions are spherical, their form factors and the potential depend  
respectively only on $q=|\vec q|$ and $D=|\vec D|$, if the latter is taken as the distance between the centres of the two distributions.
One then gets, using \eqref{eq:3}:

\begin{eqnarray}
V(D) & = & \int{\frac{d\vec q}{(2\pi)^3}\tilde V(q)}e^{-i\vec q \cdot \vec D}\nonumber \\
& = & \frac{1}{2\pi^2}\int_{0}^{\infty}{dqF_1(q)F_2(q)\frac{\sin(qD)}{qD} }\nonumber \\
& = & \frac{1}{i(2\pi)^2}\int_{-\infty}^{\infty}{\frac{dq}{qD}F_1(q)F_2(q)e^{iqD}},
\label{eq:4}
\end{eqnarray}
by extending $q$ to the negative values in the last step.

The form factor of a point-like particle is simply the value of its charge, so that $$F_{\dot{p}^{-}}(q)=-\epsilon e.$$
That of a hydrogen-like distribution in the ground state, of Bohr radius $\dot{a}_0$ and charge $+\epsilon e$, is given by
$$F_{\dot{e}^{+}}(q)=+\epsilon e \frac{16}{\left(4+\dot{a}_0^2q^2\right)^2}.$$
For the electronic distribution of the  atoms, Ref. \cite{Formfact:2006} fits the form factors to a sum of Gaussians.
These are not directly useful, as their functional form  prevents the analytic calculation of Eq. (\ref{eq:4}).
We thus refitted these form factors (to better than 2 \%) to three hydrogen-like form factors: 
\begin{equation}F_{e}(q)=-Ze\sum\limits_{i=1}^3 c_i \frac{16}{\left(4+a_i^2q^2\right)^2},\label{eq:feq}\end{equation}  with $c_3=1-c_1-
c_2$. 
The best-fit parameters for several useful elements are given in Table \ref{Table1}. 

This enables us to perform analytically the integral \eqref{eq:4} by residues. We then get for the three last terms of \eqref{eq:1}:
\begin{eqnarray}
V_{N\dot{e}^{+}}(r) & = & +\frac{Z\epsilon \alpha}{r}\left(1-\left(1+\frac{r}{\dot{a}_0}\right)e^{-2r/\dot{a}_0}\right)\label{eq:5}, \\
V_{e\dot{p}^{-}}(r) & = & +\frac{Z\epsilon \alpha}{r}\sum\limits_{i=1}^3c_i\left(1-\left(1+\frac{r}{a_i}\right)e^{-2r/a_i}\right)\label{eq:6}, 
\\
V_{e\dot{e}^{+}}(r) & = & -\frac{Z\epsilon \alpha}{r}\sum\limits_{i=1}^3c_i\left(1+\frac{1}{2}\frac{e^{-2r/a_i}S_i(r)+e^{-2r/\dot{a}_0}
T_i(r)}{U_i} \right)\label{eq:7},
\end{eqnarray}
where $S_i(r)$ and $T_i(r)$ are order-1 polynomials  in $r$ and $U_i$ is a constant depending on $\dot{a}_0$ and $a_i$, as defined in 
Appendix \ref{appendixB} where the development for \eqref{eq:7} is shown as an example. Note that in \eqref{eq:5}, we have assumed that the   
nucleus is point-like, so that we can analytically integrate. This is a good approximation: as the $N-\dot{e}^{+}$ potential is subdominant, the 
approximation changes the total 
potential  by less than 3 \% as long as $\dot{a}_0\geq 20 ~{\rm fm}$.
\begin{table}
\begin{center}
\begin{tabular}{l|c|c|c|c|c}
\hline
\hline
 \multicolumn{1}{c|}{$Z$} & $a_1$ (\AA) & $a_2$ (\AA) & $a_3$ (\AA) & $c_1$ & $c_2$ \\
\hline
\hline
 11 (sodium $1+$) & 0.04052 & 0.1961 & 0.2268 & 0.1639 & -1.1537 \\
 14 (silicon) & 0.1777 & 0.9300 & 0.001500 & 0.6295 & 0.2994 \\
 18 (argon) & 0.1082 & 0.0001260 & 0.5452 & 0.4567 & 0.02783 \\
 26 (iron) & 0.3125 & 0.05739 & 1.4872 & 0.6531 & 0.2878 \\
 32 (germanium) & 0.2157 & 0.03320 & 0.9148 & 0.6876 & 0.1834 \\
 53 (iodine 1-) & 0.9232 & 0.2728 & 0.06153 & 0.1385 & 0.4776 \\
 54 (xenon) & 0.05898 & 0.2543 & 0.7585 & 0.3755 & 0.4690 \\
 81 (thallium) & 0.01509 & 0.1199 & 0.4176 & 0.1454 & 0.5336 \\
 82 (lead) & 0.01880 & 0.1193 & 0.4171 & 0.1505 & 0.5264 \\
 \hline
 \hline
\end{tabular}
\end{center}
\caption{Best-fit parameters for the electronic form factors of several relevant atoms or ions. We consider Na$^{+}$ (sodium 1+) and I$^{-}$ 
(iodine 1-) as they are present under their ionized forms in the NaI(Tl) crystals of DAMA.}
\label{Table1}
\end{table}

\begin{figure}
\begin{center}
\includegraphics[scale=0.7]{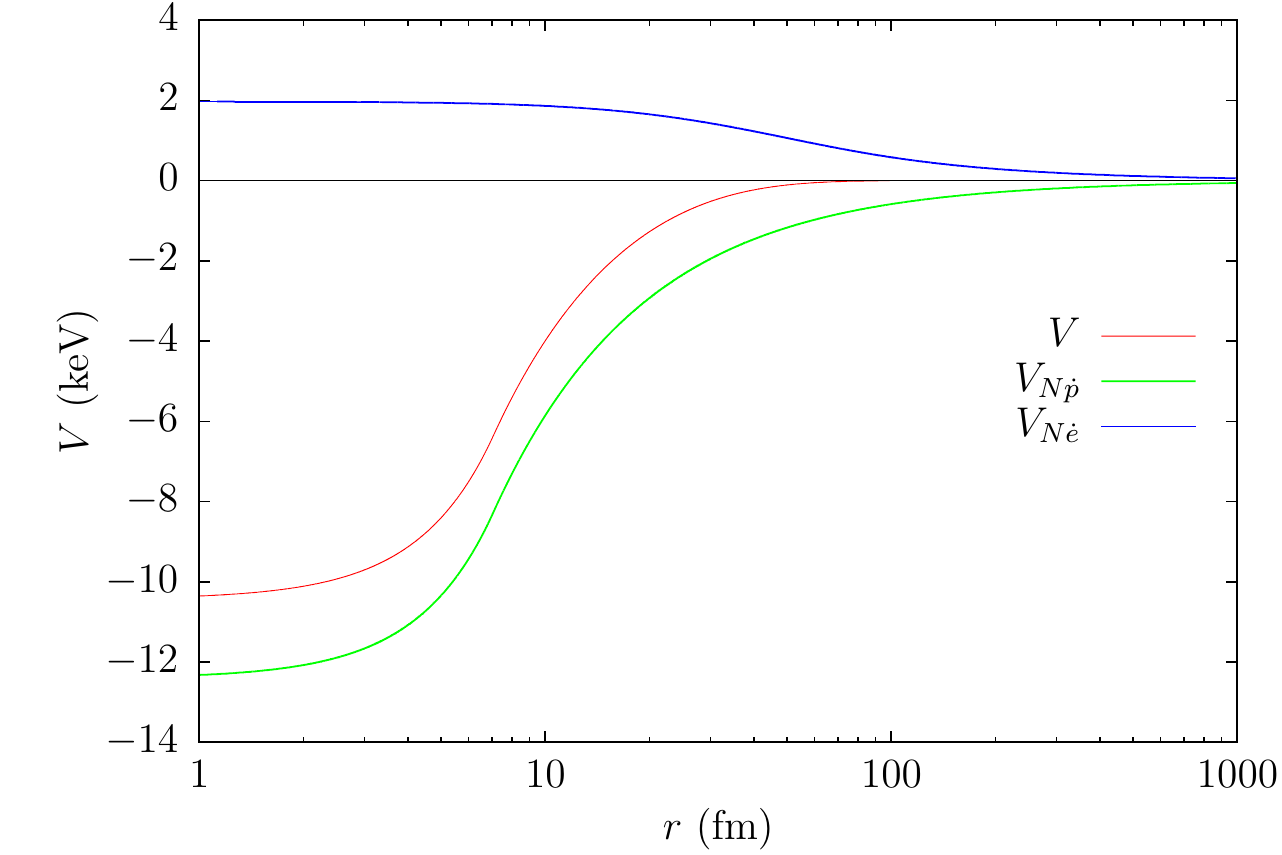}
\end{center}
\caption{Total thallium-dark atom potential $V$ (keV) (red) as a function of the radial distance $r$ for typical parameters, $
\epsilon=5~10^{-4}$ and $\dot{a}_0=30$ fm. The two dominant terms, corresponding to the nucleus-$\dot{p}^{-}$ attraction 
$V_{N\dot{p}^{-}}$  (green) and to the nucleus-$\dot{e}^{+}$ repulsion $V_{N\dot{e}^{+}}$ (blue), are also shown.}
\label{atatpotential}
\end{figure}

The atom-dark antiatom potential $V$ is therefore the sum of two attractive terms, $V_{N\dot{p}^{-}}$ and $V_{e\dot{e}^{+}}$, and two 
repulsive ones, $V_{N\dot{e}^{+}}$ and $V_{e\dot{p}^{-}}$. We show the total potential and the two dominant terms in 
Fig.~\ref{atatpotential} for a thallium atom and for the typical parameters $\epsilon=5.10^{-4}$ and $\dot{a}_0=30$ fm. 
We see that the potential is zero at large distances, 
as the two neutral atomic structures are well separated. As the electron cloud starts to merge with the dark positron one, the repulsion between 
the nucleus of the atom and the dark positron cloud that is located between the nucleus and the dark 
antiproton induces a very small potential barrier, within the thickness of the lines in the figure. After the nucleus  
 enters the dark positron cloud, the potential becomes attractive. For $r\leq \dot{a}_0$, the interaction between the
nucleus and the dark antiproton $V_{N\dot{p}^{-}}$ dominates. 

We numerically solve  the radial Schr\"odinger equation for the wave function for the atom - dark antiatom relative motion with the potential 
$V(r)$ of Eq.~ \ref{eq:1} to find the bound states. As usual,
we take the wave function to be $\psi(\vec r)={u_{n\ell} (r)\over r}Y_{\ell m}(\theta,\phi)$. For a state of a given $\ell$, one 
knows that 
for $r\rightarrow 0$, $u_{n\ell} (r)\rightarrow N_0 r^{\ell+1}$. Asymptotically, for $r\rightarrow\infty$, one knows that $u_{n\ell} (r)\rightarrow N_
\infty \exp(-\sqrt{-2E\mu} r)$ with $\mu$ the reduced mass of the atom - dark antiatom system and $E<0$ the energy. The energy equals the potential at the turning 
point, which we shall call $r=a_t$, and it defines two regions. To fix the constants $N_0$ and $N_\infty$, we calculate the wave function 
numerically, using a Runge-Kutta-Dormand-Prince (RKDP \cite{RKDP}) method, starting forward from 0 to $a_t$ for the small-$r$ region, and backwards from 
$r_{max}=5 a_t$ to $a_t$ for the large-$r$ region. At $a_t$, we impose that $u_{n\ell}(r)$ is continuous, and we scan the value of the binding
energy until the derivatives at $a_t$ are also symmetric, which gives us the energy that is solution. We finally normalize the wave function to unity.
 Note that the method has been checked to  give results  consistent with those obtained through the WKB approximation.

\begin{figure}
\begin{center}
\includegraphics[scale=0.7]{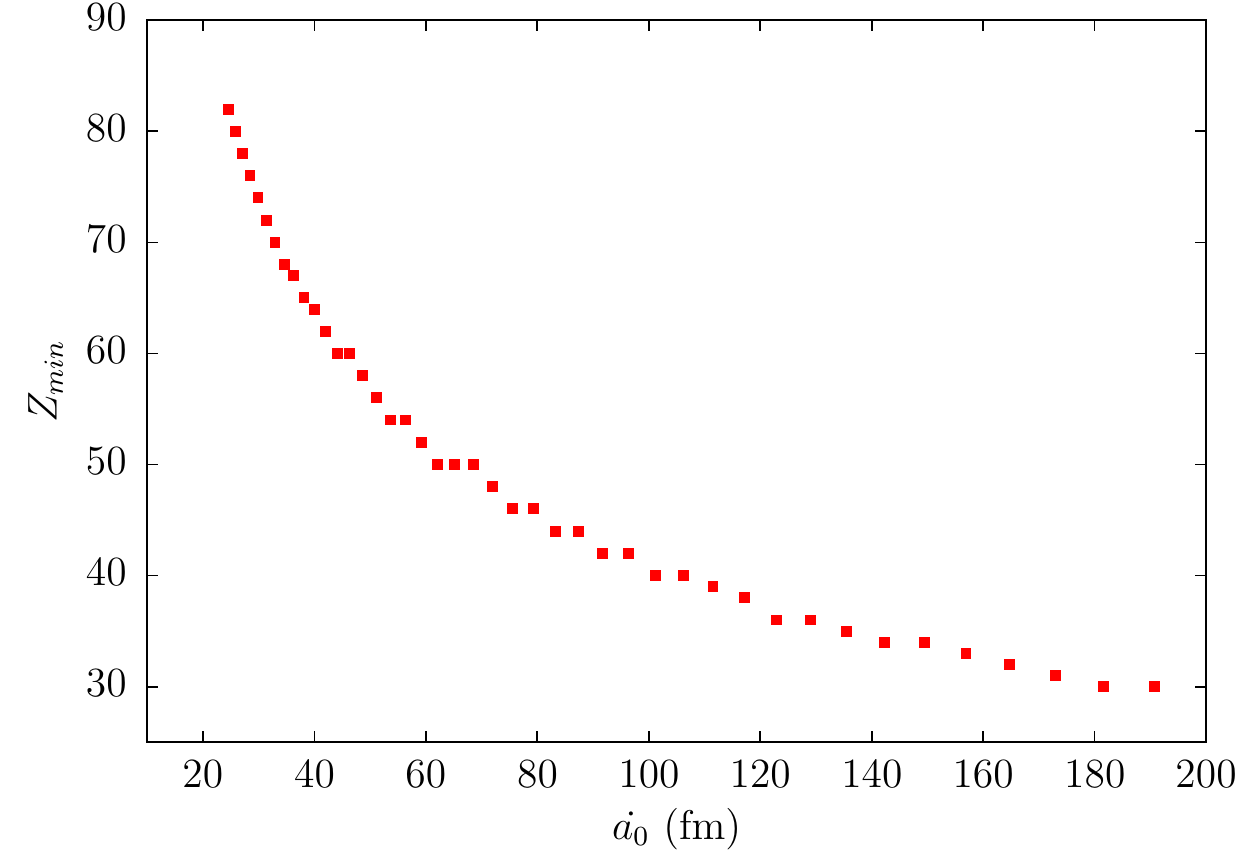}
\end{center}
\caption{Atomic number $Z_{min}$ from which at least one $p$ bound state exists, and hence radiative capture is possible, as a function of the 
Bohr 
radius $\dot{a}_0$ (fm) of the dark antiatom, for $m=1000$ GeV and $\epsilon=5.10^{-4}$.}
\label{Zmina0p}
\end{figure}

We will see in Section \ref{secDAMArate} that the radiative capture of the dark atoms by visible ones requires the existence of at least one $p$ 
state in the potential $V(r)$. For $\dot{a}_0$ between $20$ and $200$ fm  and for $m=1000$ GeV and $\epsilon=5.10^{-4}$, we sought the 
first stable element $Z_{min}$ for which a $p$ bound state appears, and we show the result in Figure \ref{Zmina0p}. We see that $Z_{min}=74$ 
(tungsten) for $\dot{a}_0=30$ fm, showing that for compact dark antiatoms, binding is possible only with very heavy elements.
\section{Thermalization and DAMA event rate}\label{secDAMArate}
Because of the orbital motion of the Earth around the Sun and of the Sun around the Galactic center, dark antiatoms hit the surface of the Earth 
continuously. When they penetrate the crust, dark antiatoms undergo collisions with terrestrial atoms. As bound states can form only with very 
heavy elements and not with the rather light ones constituting the crust, these collisions are purely elastic. Only lead is sufficiently heavy 
($Z=82$) and abundant ($10$ ppm) to be considered and its effect, which is in fact negligible, will be discussed in Section \ref{AnElem}. Due to 
the repeated collisions, the dark atoms deposit their energy in terrestrial matter until they thermalize completely. This happens before the dark 
atoms have reached the underground detectors, located at about $1$ km deep, and results in a descending cloud of dark antiatoms driven by 
gravity towards the center of the Earth. When that cloud enters a detector, the thermal energies are too low to give rise to nuclear recoils, but 
the emission of photons caused by the radiative capture of the dark antiatoms by atoms of the active medium (thallium for DAMA) 
produces the signal. The constraints from the other experiments are avoided by the absence of bound states with their constituent atoms.

To simplify the problem, we use the result of ref. \cite{Sigurdson:2004zp}: it is enough to approximate the crust of the Earth as
being made of pure Si, with atomic and mass numbers $Z=14$ and $A=28$. In the Born 
approximation, the differential elastic cross section $\frac{d\sigma}{d\Omega}$ is related to the Fourier transform of the interaction potential 
\eqref{eq:1} and so to the form factors of the silicon and dark antiatoms as seen in \eqref{eq:3}:
\begin{eqnarray}
\frac{d\sigma}{d\Omega} & = & \frac{\mu^2_{Si\dot{p}^{-}}}{4\pi^2}\left|\int d\vec{r}~V(r) e^{-i\vec{q}\cdot \vec{r}}\right|^2\nonumber 
\\
& = & \frac{\mu^2_{Si\dot{p}^{-}}}{4\pi^2}\frac{(F_{N}(q)+F_e(q))^2(F_{\dot{p}^{-}}(q)+F_{\dot{e}^{+}}(q))^2}{q^4},
\end{eqnarray}
where $\mu_{Si\dot{p}^{-}}=m_{Si}m/(m_{Si}+m)$ is the reduced mass of the silicon-dark antiatom system, $m_{Si}$ is the mass of a silicon 
nucleus, $m$ that of the dark antiatom, $\vec{q}$ is the transferred momentum. Here we take a point-like silicon nucleus, {\it i.e.}
 $F_N(q)=14e$, which is  a very good approximation since $qR\ll1$ for the whole thermalization process. The parameters for the electronic 
 form factor $F_e(q)$ of a silicon atom are given in Table \ref{Table1}. In the center-of-mass frame, $q=2p\sin\theta/2$, where $p$ is the 
 initial momentum and $\theta$ is the deflection angle with respect to the collisional axis.

At each collision with an atom at rest in the crust, a dark atom loses an energy $\Delta K=\frac{p^2(\cos\theta -1)}{m_{Si}}$ in the frame of the Earth, which results in an energy loss per unit length:
\begin{equation}
\frac{dE}{dx}=n_{Si}\int_{\Omega} \Delta K \left(\frac{d\sigma}{d\Omega}\right) d\Omega,
\end{equation}
if one assumes that the path of a dark antiatom through terrestrial matter is linear. 
This is valid here since $m\gg m_{Si}$. $n_{Si}\approx 5~10^{22}$ cm$^{-3}$ is the number density of atoms in the terrestrial crust.

Therefore, the penetration length $x$  of a dark antiatom under the surface is given by:

\begin{equation}
x=\int_{E_{th}}^{E_{0}} \frac{dE}{|dE/dx|},
\end{equation}
where $E_0$ and $E_{th}=\frac{3}{2} T_{crust}$ are respectively its initial kinetic energy and final thermal energy, with $T_{crust}\simeq 
300$ K. We require that $x<1$ km. After that distance the thermalized dark antiatoms start to drift down towards the center of the Earth 
because of gravity, with a drift a velocity $$V_d=g/n_{Si}\left<\sigma v\right>,$$ where $g=980$ cm/s$^2$ is the acceleration of gravity, $
\sigma$ is the total elastic cross section and $v$ is the relative velocity. The product $\sigma v$ is averaged over a Maxwell-Boltzmann velocity 
distribution of temperature $T_{crust}$ for both the silicon atoms and the dark antiatoms. Typically, $V_d\approx 10$ cm/s, 
so that a change in 
the incident flux at the surface is felt at most $3$ hours later for a detector located at a depth of $1$ km.

To get the number density $n$ of dark antiatoms in the detector, we balance the incident flux at the surface of the Earth with the down-drifting 
thermalized flux: $$\frac{n_0}{4}|\vec{V}_{\odot} + \vec{V}_{\oplus}|=nV_d,$$ where $n_0$ (cm$^{-3}$)$=0.3f/m$ (GeV) is the local number density of dark atoms and $\vec{V}_{\odot},\vec{V}_{\oplus}$ are respectively the orbital velocity of the sun around the center of the galaxy and the orbital velocity of the Earth around the sun. The number density of dark atoms near the sun is obtained by considering a fraction $f\in]0,1]$ of the local dark matter density of $0.3$ GeV/cm$^{-3}$, as explained in Section \ref{darkSect}. Due to the periodic orbital motion of the Earth around the sun, the norm $|\vec{V}_{\odot} + \vec{V}_{\oplus}|$ is annually modulated and so is $n$:

\begin{equation}
n(t)=N^0+N^m\cos(\omega (t-t_0)),\label{eq:11}
\end{equation}
where $\omega=2\pi/T_{orb}$ is the angular frequency of the orbital motion of the Earth, with a period $T_{orb}=1$ yr, $t_0\simeq$ June 
$2$ is the day of the year when $\vec{V}_{\odot}$ and $\vec{V}_{\oplus}$ are best aligned and where the constant and modulated parts 
$N^0$ and $N^m$ are given by:

\begin{eqnarray}
N^0 & = & \frac{n_0n_{Si}\left<\sigma v\right>}{4g}V_{\odot},\label{eq:12} \\
N^m & = & \frac{n_0n_{Si}\left<\sigma v\right>}{4g}V_{\oplus}\cos\gamma,\label{eq:13}
\end{eqnarray}
where $V_{\odot}=|\vec{V}_{\odot}|=220$ km/s, $V_{\oplus}=|\vec{V}_{\oplus}|=29.5$ km/s and $\gamma\simeq 60^{\circ}$ is the angle 
between the orbital plane of the Earth and the Galactic plane.

Since there are no bound states with the sodium ($Z=11$) and iodine ($Z=53$) components of the DAMA detectors, the signal is entirely 
due to the thallium dopant, present at the $10^{-3}$ level, i.e. with a number density $n_{Tl}=10^{-3}n_{NaI}$, where $n_{NaI}$ is the number 
density of the sodium iodide crystal. Thermal collisions within the detector between dark atoms and thallium give rise to  a rate per unit volume for 
 bound-state-formation:
\begin{equation}
\Gamma(t)=n_{Tl}~n(t)\left<\sigma_{capt}v\right>=\Gamma^0+\Gamma^m\cos(\omega (t-t_0)),
\label{eq:14}
\end{equation}
where $\sigma_{capt}$ is the radiative capture cross section and $v$ the relative velocity. This rate is also modulated due to \eqref{eq:11}.
The product $\sigma_{capt}v$ is thermally averaged over two Maxwell-Boltzmann velocity distributions , for thallium and the dark atoms that have entered the detector. Both distributions are for temperature $T=300$ K, as DAMA operates at room temperature.

Because of the spin independence of the interaction, magnetic dipole transitions are forbidden and the dominant type of capture is therefore 
electric dipole (E1). As the dark atom is at low energy $E$ ($>0$) in the center-of-mass frame of the atom-dark atom system, the $s$ wave is 
the dominant term in the expansion of the incident plane wave into spherical waves. Therefore, the final state of energy $E_p$ ($<0$) has to be 
a $p$ state due to the selection rules of E1 transitions and the capture causes the emission of a photon of energy $|E-E_p|\simeq |E_p|$.

We solve the radial Schr\"odinger equation for the energy $E$ by the RKDP method \cite{RKDP} to obtain the radial part $R(r)$ of the initial 
diffusion eigenfunction and with the method discussed in Section~\ref{Binding} to get $E_p$ and the radial part $R_p(r)$ of the wave function of 
the final bound state. The capture cross section is then obtained by computing the matrix element $\mathcal{M}=\int_{0}^{\infty} 
rR_p(r)R(r)r^2dr$ of the dipole operator between these two states and has been shown \cite{Wallemacq:2013hsa} to be:
\begin{equation}
\sigma_{capt}=\frac{32\pi^2 Z^2 \alpha}{3\sqrt 2}\left( \frac{m}{m+m_{Tl}} \right)^2 \frac{1}{\sqrt{\mu_{Tl\dot{p}^{-}}}} \frac{\left( E-
E_p\right)^3}{E^{3/2}}\mathcal{M}^2,
\end{equation}
where $m_{Tl}$ is the mass of thallium, for which $Z=81$, and $\mu_{Tl\dot{p}^{-}}$ the reduced mass of the thallium-dark antiatom system. 
After this capture, the system de-excites to an $s$ state of energy $E_s$ via a second E1 transition, which causes the emission of a second 
photon of energy $E_{p}-E_{s}$. To avoid the observation of double-hit events in the detector, we require that the first emitted photon be not 
seen, {\it i.e.} that its energy be below the threshold of the experiment at $2$ keV. The signal is thus due to the second transition, which in principle can give rise to a spectrum as there are in general 
several $p$ and $s$ states in the well leading to several transitions of different energies. For simplicity, we shall assume here that the signal is 
due to the dominant transition, from the lowest $p$ state, of energy $E_p=E_1$, to the ground state, of energy $E_s=E_0$. As the hits 
observed by DAMA lie between $2$ and $6$ keV, we require $$2~{\rm keV} \leq E_1-E_0\leq 6~\rm  keV,$$
while the condition related to the double-hit events gives $$|E_1|\leq 2~\rm keV.$$

Finally, using \eqref{eq:14}, passing to the center-of-mass and relative velocities $\vec{v}_{CM}$ and $\vec{v}$, performing the integral over 
the center-of-mass variables before inserting \eqref{eq:11} and expressing the event rate in counts per day and per kilogram of detector 
(cpd/kg), we get for the constant and modulated parts $\Gamma^0$ and $\Gamma^m$:

\begin{eqnarray}
\Gamma^0 & = & CN^0\int_{0}^{\infty}\sigma_{capt}(E)Ee^{-E/T}dE, \\
\Gamma^m & = & CN^m\int_{0}^{\infty}\sigma_{capt}(E)Ee^{-E/T}dE, \\
C & = & \frac{9.71~10^{11}}{M_{NaI}\sqrt{\mu_{Tl\dot{p}^{-}}}(T)^{3/2}},
\end{eqnarray}
where  $M_{NaI}=150$ g/mol is the molar mass of NaI, and where $N^0$ and $N^m$ given by \eqref{eq:12} and \eqref{eq:13} have to be 
expressed in cm$^{-3}$, $\sigma_{capt}$ in GeV$^{-2}$ and $\mu_{Tl\dot{p}^{-}}$, $T$, $E$ in GeV in order to get the event rate in cpd/kg. 

It should be noted here that the photons emitted during the capture process produce electron recoils instead of nuclear recoils in the case of 
WIMPs. However, as DAMA does not discriminate between the two types of recoils, its results can be directly reinterpreted via the capture 
described here.

\subsection{Results}
\begin{figure}
\begin{minipage}{0.49\linewidth}
\centerline{\includegraphics[width=1\linewidth]{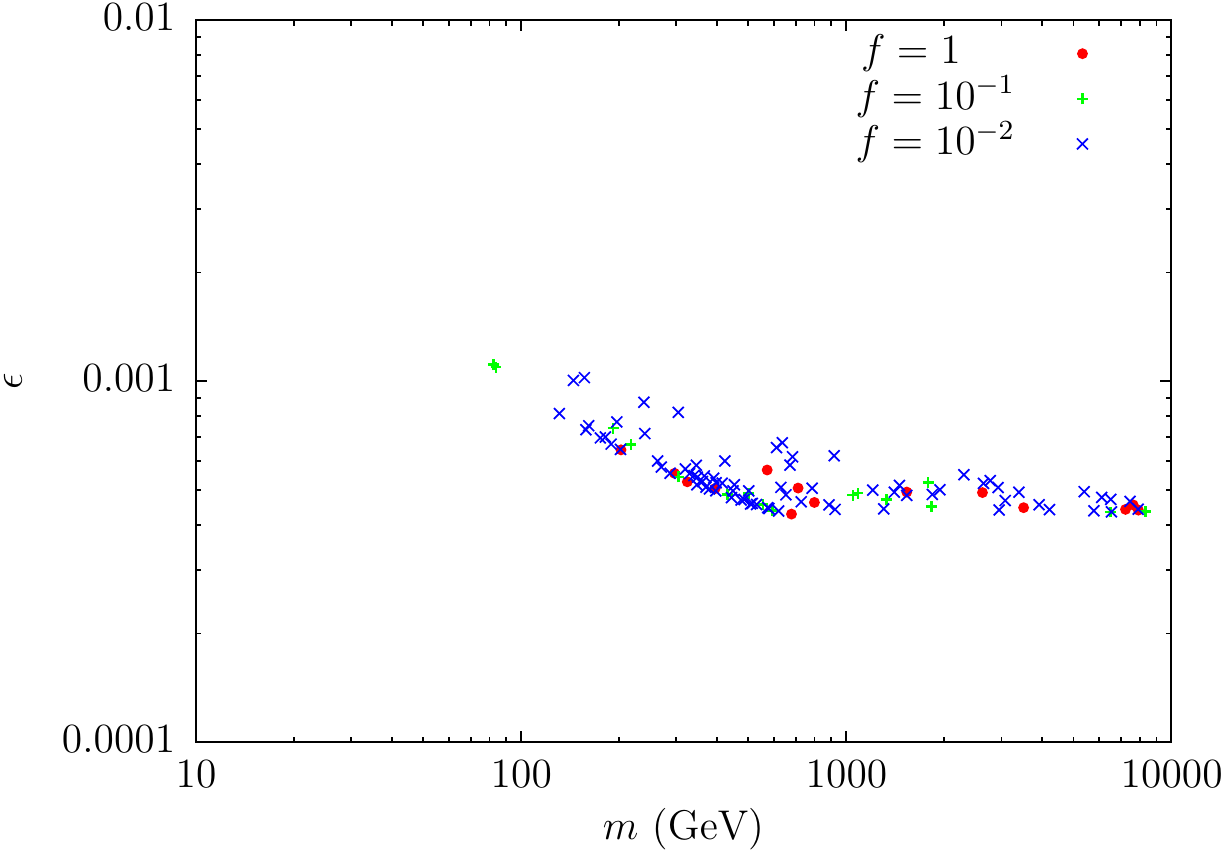}}
\end{minipage}
\begin{minipage}{0.49\linewidth}
\centerline{\includegraphics[width=1\linewidth]{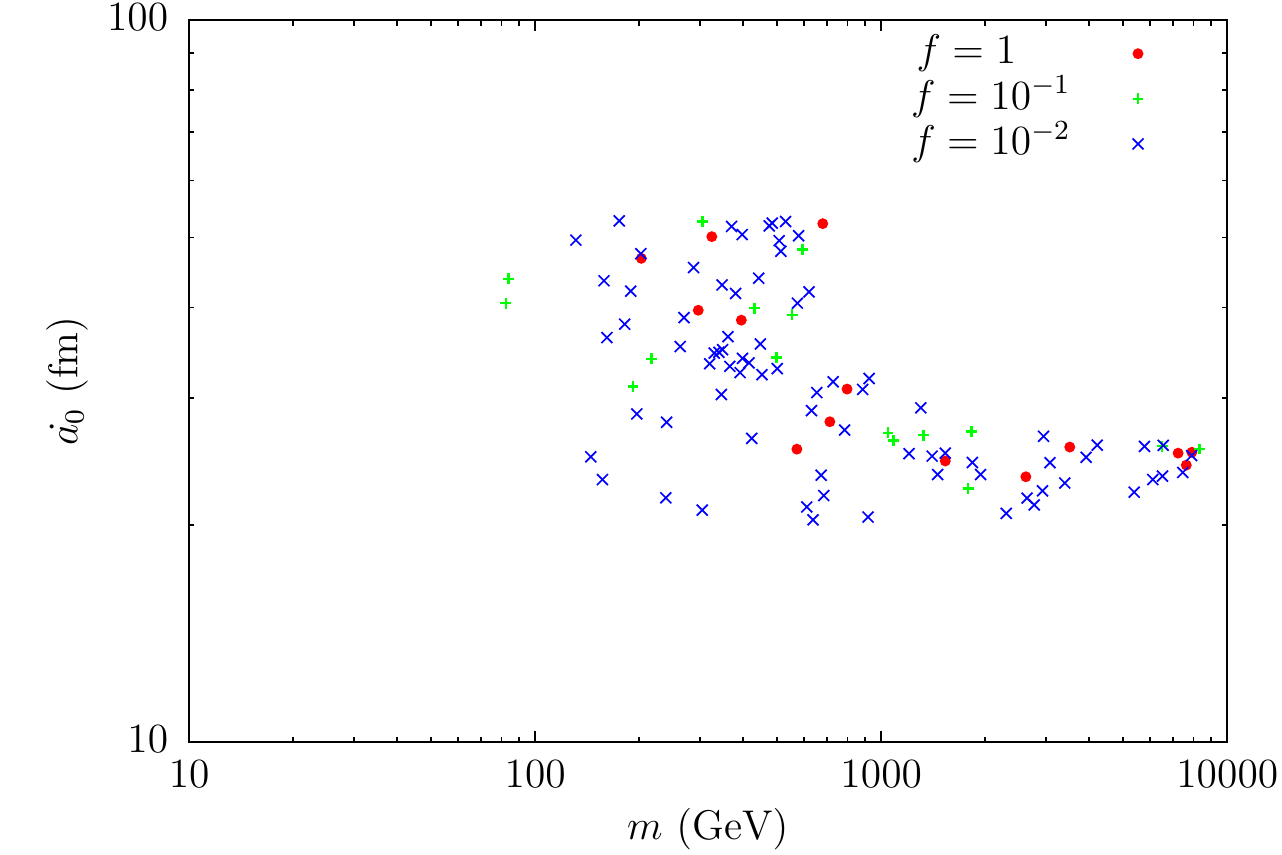}}
\end{minipage}
\begin{center}
\includegraphics[scale=0.6]{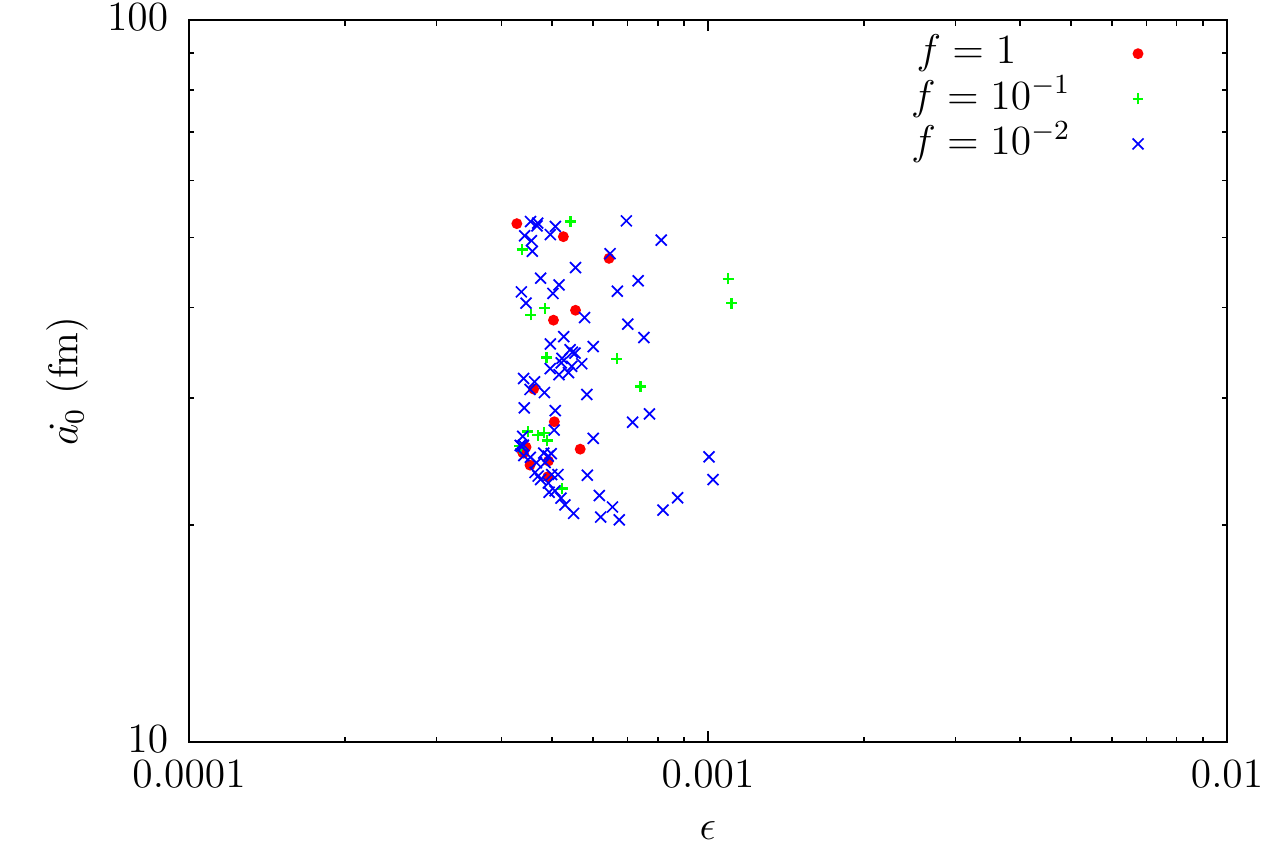}
\end{center}
\caption{Regions of the $3$-dimensional parameter space that reproduce the DAMA/LIBRA results with thallium at the $2\sigma$ level, projected on the $(m,\epsilon )$-plane (top left), on the $(m,\dot{a}_0 )$-plane (top right) and on the $(\epsilon,\dot{a}_0)$-plane (bottom center). The models corresponding to $f=1,10^{-1}$ and $10^{-2}$ are shown respectively with red dots, green crosses and blue crosses.}
\label{regions}
\end{figure}

DAMA/LIBRA observes an annually modulated rate $\Gamma^{m}_{DAMA}=(0.0448\pm 0.0048)$ cpd/kg between 2 and 6 keV. To reproduce 
it at the $2\sigma$ level, we  randomly sampled the $(m,~\epsilon,~\dot{a}_0)$ space for $1$ GeV $\leq m\leq 50$ TeV, $10^{-5}\leq 
\epsilon \leq 10^{-2}$ and $10$ fm $\leq \dot{a}_0\leq 1$~\AA~and generated several millions of models. For each of them, we required the 
existence of at least one $p$ state with thallium, an energy for the first E1 transition below the DAMA threshold, a second E1 transition in 
the observed energy range, with the correct rate, thermalization before $1$ km and the absence of bound states with sodium and iodine. The 
parameters of the successful models are shown in Figure \ref{regions} for different fractions of dark antiatoms, $f=1,10^{-1}$ and $10^{-2}$. 

The models that fulfil all the conditions are characterized by $100$ GeV $\leq m\leq 10$ TeV, $4.10^{-4}\leq \epsilon \leq 10^{-3}$ and $20$ 
fm $\leq \dot{a}_0\leq 50$ fm. The cut-off on $m$ around $10$ TeV is due to the requirement of thermalization before $1$ km while the range 
of values of $\epsilon$ is a direct consequence of the existence of energy levels in the keV region, such that the transition lies between $2$ and 
$6$ keV. For $\dot{a}_0$, rather small values are needed to decrease the usually large atomic capture cross sections and reach the event rate 
of DAMA/LIBRA. As larger capture cross sections are easily obtained, we expect that, in the case of models where this is compensated by a 
smaller number of incoming particles ($f<1$), the regions in the parameter space are more extended and denser, which we verify in Figure 
\ref{regions} for $f=10^{-1}$ and $f=10^{-2}$.

We also checked that there are no bound states with xenon, and hence with the elements of atomic numbers $Z\leq 54$, so that the negative results of CDMS-II/Ge and superCDMS (both detectors made of germanium) and XENON100 and LUX (both detectors made of xenon) can be naturally explained.

\subsection{Absorption by lead and anomalous elements}\label{AnElem}
Models involving capture will always lead to anomalous heavy isotopes, to a corresponding decrease in the flux of dark matter, and to
an eventual signal originating from the shielding of the detector. We shall now examine these effects, and show that they do not pose a problem, 
but may offer a way to test this model. In the following discussion, we concentrate on typical values of the parameters as in Fig.~2: $m=1000$ 
GeV, $\epsilon=5.10^{-4}$ and $\dot{a}_0=30$ fm.
\subsubsection{Absorption in the terrestrial crust}

We have seen in Section \ref{Binding} that the dark antiatoms bind only to very heavy elements. This can be seen in Figure \ref{Zmina0p} for the 
interesting values of $\dot{a}_0$. For the parameters that we chose, the first element that has bound states is tungsten, with $Z=74$. 
In the terrestrial crust, the only element beyond tungsten that is sufficiently abundant to be considered is lead, with $Z=82$, which is present at 
the level of about $10$ ppm. We shall first check that the absorption by this element on the way down through terrestrial matter does not 
significantly reduce  the available dark matter flux.

Let us consider the  case of a  dark matter column above the DAMA/LIBRA detector with the same section. As the capture cross section is 
maximal at low energy, an upper bound on absorption
can be obtained by taking all the dark atoms at the thermal energy, as if they thermalized as soon as they touched the surface of the Earth. 

To estimate the reduction of the initial flux, we first need the flux of dark antiatoms in the column. The available flux at the surface of the Earth, 
for a fraction $f$ and dark atoms of mass $m=1000$ GeV, is given by $\phi_i=n_0V_{\odot}/4$ which must be equal to the thermalized down 
drifting flux $\phi_d(0)=n(0)V_d$, so that $$n(0)=165f  {\rm cm}^{-3},$$ if we take $V_d=10$ cm/s. $n(0)$ is here the 
available number density of thermalized dark atoms just below the surface.

The capture cross section can also be extracted directly from the data.
DAMA/LIBRA observes a modulation amplitude of $0.0448$ cpd/kg, which corresponds to a constant part of the rate $\Gamma^0_{DAMA}$ of 
$0.672$ cpd/kg since $\Gamma^0/\Gamma^m\simeq15$, or to $2.85~10^{-8}$ counts/s/cm$^3$ using the density of sodium iodide $
\rho_{NaI}=3.67$ g/cm$^3$. Thallium is present at the $10^{-3}$ level in the detector, i.e. $n_{Tl}=10^{-3}n_{NaI}=1.5~10^{19}$ cm$^{-3}
$. The capture rate can be roughly expressed as $\Gamma^0_{DAMA}=n_{Tl}~n(0)\sigma_{capt}V_t$, where $V_t=\sqrt{\frac{8T}{\pi 
\mu_{Tl\dot{p}^{-}}}}= 1.7~10^4$ cm/s is the mean relative velocity between dark antiatoms and thallium in the detector at temperature 
$T=300$ K. We can therefore access the capture cross section of the dark antiatoms by thallium at thermal energy: $$\sigma_{capt}
=6.9~10^{-34}/f {\rm cm}^2.$$

Since thallium and lead are very close elements ($Z=81$ and $Z=82$ and similar masses), we will assume that their 
capture cross sections are the same. $\sigma_{capt}$ will therefore be used along the whole thermalized column from 
the surface to the detector to estimate the absorption by lead. The variation of the down drifting dark matter flux $
\phi_d=n(x)V_d$ at depth $x$ is given by $$V_d\frac{dn(x)}{dx}=-n_{Pb}n(x)V_t \sigma_{capt},$$ from which 
we obtain $n(x)=n(0)\exp(-n_{Pb}\left(V_t/V_d\right)\sigma_{capt}x)$, where $n_{Pb}$ is the number density of lead in the terrestrial crust. Since its abundance is $10$ ppm, $n_{Pb}\sim 10^{17}$ cm$^{-3}$. As the DAMA/LIBRA detector is located at a depth $L=1$ km, we finally get the following estimate for the relative variation of the dark matter flux:
\begin{equation}
\left|\frac{\phi_d(L)-\phi_d(0)}{\phi_d(0)}\right|=\left|\frac{n(L)-n(0)}{n(0)}\right|=1-e^{-n_{Pb}\left(V_t/V_d\right)\sigma_{capt}L}\simeq n_{Pb}\left(V_t/V_d\right)\sigma_{capt}L=1.17~10^{-8}/f.
\label{eq:19}
\end{equation}
Clearly, the absorption by the Earth is negligible:
for the values of $f$ that we considered, it does not exceed $1.17~10^{-6}$, and this is why we considered elastic collisions only in the
thermalization process.

\subsection{Absorption in the shield of the detector}

\begin{figure}
\begin{center}
\includegraphics[scale=0.7]{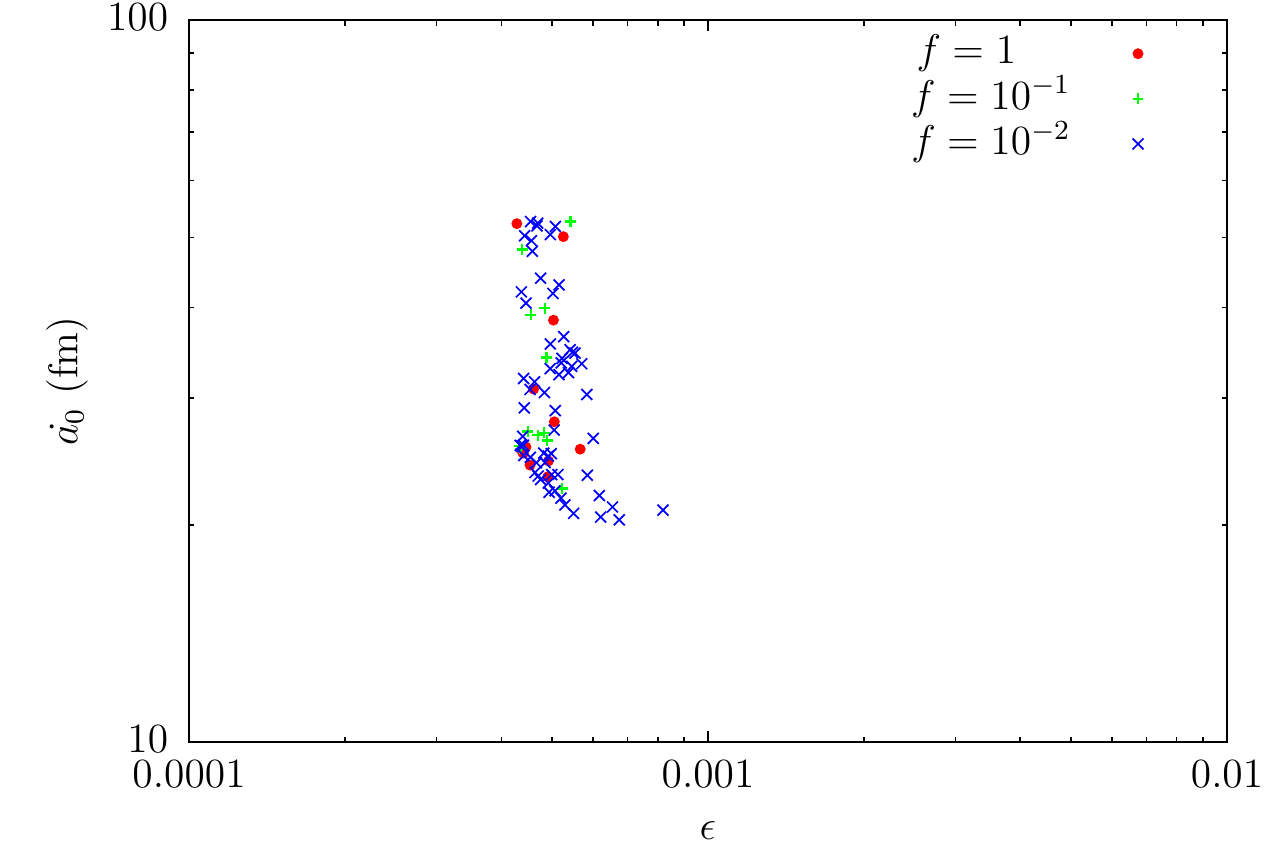}
\end{center}
\caption{Region of the $3$-dimensional parameter space that reproduces the DAMA/LIBRA results with thallium at the $2\sigma$ level, 
projected on the $(\epsilon,\dot{a}_0)$-plane, with the account for the constraint on $m$ from anomalously heavy isotopes of gold: $m>300$ 
GeV. The models corresponding to $f=1,10^{-1}$ and $10^{-2}$ are shown respectively with red dots, green crosses and blue crosses.}
\label{regions_gold}
\end{figure} 

Another source of lead, which could be more problematic, is the shield of the detector. Indeed, underground dark matter detectors are usually 
shielded with several layers of lead in order to isolate them from cosmic rays or environmental radioactivity.

A column of rock of height $1$ km has a lead density of $3.5$ g/cm$^2$ while the layer of $15$ cm of lead on top 
of the DAMA/LIBRA apparatus has a density of $170.25$ g/cm$^2$. From \eqref{eq:19}, this corresponds to a relative correction of the event 
rate of $\frac{170.25}{3.5}\times1.17~10^{-8}/f=5.7~10^{-7}/f$, which is still negligible with respect to the experimental uncertainty, even 
for the smallest considered value of $f$. As the experimental error on the rate of DAMA/LIBRA is of the order of $10\%$ of the central 
measured value, absorption by lead in the shield could become important for $f\lsim 10^{-5}$.

Note that there are other materials constituting the shield of the detector, such as copper or cadmium, but these are lighter and therefore do not 
contribute to the reduction of the flux since they do
not have any bound states with the dark antiatoms..

\subsection{Limits from superheavy elements}\label{super}

Since the birth of the Earth, dark antiatoms have been binding to the heavy stable nuclei that constitute it.
($Z\geq 74$ for our typical model), resulting in the accumulation of anomalous superheavy isotopes of known elements.
To estimate their abundance today, we can use the formation rate of superheavy isotopes of thallium from DAMA/LIBRA: dark atoms bind to 
thallium at a rate of $0.672$ counts per day and per kilogram of NaI, i.e. $0.672\times \frac{150}{1000}=0.1$ counts per day and per mole of 
NaI. As in $1$ mole of components of the detector, $0.001$ is made of thallium, this gives $0.1\times1000=100$ superheavy isotopes of 
thallium formed per day and per mole of thallium. Therefore, over the whole history of the Earth, which is $4.5~10^{9}$ years old, 
$100\times365\times4.5~10^{9}=1.64~10^{14}$ anomalous isotopes have been produced per mole, which gives an abundance of $
\frac{1.64~10^{14}}{6.02~10^{23}}=2.7~10^{-10}$.

In Ref. \cite{PhysRevD.65.072003}, limits on the abundance of superheavy gold have been obtained by analysing several terrestrial samples in an 
accelerator mass spectrometer. These limits depend on the mass $M$ of the superheavy isotope and get weaker as the mass increases. Since 
thallium and gold are close elements ($Z=81$ and $Z=79$ respectively), it is reasonable to use the abundance of superheavy thallium as an 
approximation for gold. For that abundance,  $M>500$ GeV is required. As $M=m_{Au}+m$, where $m_{Au}\simeq 200$ GeV is the mass of a 
gold atom, this gives $m>300$ GeV from the search for anomalously heavy isotopes of gold.

Therefore, the two upper subfigures of Figure \ref{regions} should be viewed with a lower limit on $m$ at $300$ GeV, while the bottom 
subfigure of the $(\epsilon ,\dot{a}_0)$-plane, taking into account this new constraint, becomes Figure \ref{regions_gold}.

\section{Conclusion}
We have proposed here a very simple model that can explain why only DAMA observes a signal. The model assumes that
a subdominant form of dark matter is made of dark antiatoms. It depends on three parameters:
the mass $m$ of the dark antiatom,  its  Bohr radius $ \dot{a}_{0}$, and the millicharge $\epsilon e$ of its antinucleus.
The annual modulation of DAMA can be reproduced via the binding to the thallium dopant of the NaI scintillator for
$300$ GeV $\leq m\leq 10$ TeV, $\epsilon \sim 5.10^{-4}$ and $20$ fm $\leq \dot{a}_{0}\leq 50$ fm.
There is no bound state with the sodium or the iodine components of the detector, so that the signal is entirely due to thallium. 
The constraints from CDMS-II/Ge, superCDMS, XENON100 and LUX disappear because their  constituent 
nuclei do not bind to the dark antiatoms and because the thermal energies of the dark antiatoms are insufficient 
to produce detectable nuclear recoils.

The model is easily falsifiable, as it lies on the edge of the existing limits for superheavy anomalous elements \cite{PhysRevD.65.072003}: the 
lower bound on $m$ coming from the limits on the terrestrial abundances of superheavy isotopes of known elements. 
Furthermore,  the addition of a heavy isotope in the active part of any direct detection experiment would dramatically increase the signal
or the electronic noise.

Note that this model also implies some X-ray emission from the center of galaxies. The dark antiatoms considered here have small 
sizes $\dot{a}_0$, about $1000$ times smaller than that of  hydrogen. If we assume that $\dot{\alpha}\simeq \alpha =1/137$, then we 
expect their binding energy $\dot{E}_I=-\frac{1}{2}\frac{\dot{\alpha}}
{\dot{a}_0}$ to be shifted from the eV region to the keV region.  Collisions between dark antiatoms in the central regions of dark halos, where 
the velocities and densities of dark matter are expected to be higher, will lead to the excitation of the dark atoms, and a fraction of them will 
de-excite through the emission of  photons, as the dark current has a millicharge $\epsilon e$. This will result in the emission of a weak thin line 
 in the X-ray region. This is reminiscent of the recently observed 3.5 keV line \cite{Boyarsky:2014jta,Bulbul:2014sua}. However, 
a rough estimate of the emitted flux indicates that is is at least four times smaller than reported one. 
We plan to come back to this question in a future publication.

\section*{Acknowledgements}
We are grateful to G. Rauw for useful information concerning X-ray astronomy. This work was supported by the Fonds de la Recherche Scientifique - FNRS under grant 4.4501.05. and one of us (Q.W.) is supported by the Fonds de la Recherche Scientifique - FNRS as a research Fellow.

\appendix
\makeatletter
\def\@seccntformat#1{Appendix~\csname the#1\endcsname:\quad}
\makeatother
\section{Form factors}\label{appendixA}
Let $\rho_1(\vec x)$ and $\rho_2(\vec y)$ be two charge distributions localized by $\vec x$ and $\vec y$
from two arbitrary reference points. We can parametrize the potential energy of the electrostatic interaction 
between the two distributions, or the electrostatic interaction potential $V$, via the vector $\vec D$ that 
joins the two reference points, although it is independent on the choice of the latter. We have:

\begin{equation}
V(\vec D)=\frac{1}{4\pi}\int d\vec x d\vec y \frac{\rho_1(\vec x)\rho_2(\vec y)}{|\vec D + \vec y - \vec x|}
\end{equation}
We can then write $\rho_1(\vec x)$ and $\rho_2(\vec y)$ in terms of their Fourier transforms $\tilde{\rho}_1(\vec k)$ and $\tilde{\rho}_2(\vec l)$ and  use  the properties of the $\delta$-distribution, to get:

\begin{eqnarray}
V(\vec D) & = & \frac{1}{4\pi}\int\frac{d \vec k d\vec l}{(2\pi)^6}\int d\vec x d\vec y d\vec z \frac{\tilde{\rho}_1(\vec k)\tilde{\rho}_2(\vec l)}{|\vec z|}e^{-i\vec k \cdot \vec x}e^{-i\vec l \cdot \vec y}\delta(\vec z - \vec D - \vec y + \vec x)\nonumber \\
& = & \int\frac{d \vec k d\vec l d\vec q}{(2\pi)^9}\int d\vec x d\vec y d\vec z \frac{\tilde{\rho}_1(\vec k)\tilde{\rho}_2(\vec l)}{q^2}e^{-i\vec k \cdot \vec x}e^{-i\vec l \cdot \vec y}e^{-i\vec q \cdot \vec z}\delta(\vec z - \vec D - \vec y + \vec x)\nonumber,
\end{eqnarray}
where we have used the fact that the Fourier transform of $\frac{1}{|\vec z|}$ is $\frac{4\pi}{q^2}$ at the second line. Performing the integral over $\vec z$ and rearranging the exponentials leads to:

\begin{eqnarray}
V(\vec D) & = & \int\frac{d \vec k d\vec l d\vec q}{(2\pi)^9}\int d\vec x d\vec y \frac{\tilde{\rho}_1(\vec k)\tilde{\rho}_2(\vec l)}{q^2}e^{-i\vec k \cdot \vec x}e^{-i\vec l \cdot \vec y}e^{i\vec q \cdot \vec x}e^{-i\vec q \cdot \vec y}e^{-i\vec q \cdot \vec D}\nonumber \\
& = & \int\frac{d \vec k d\vec l d\vec q}{(2\pi)^9}\int d\vec x d\vec y \frac{\tilde{\rho}_1(\vec k)\tilde{\rho}_2(\vec l)}{q^2}e^{-i(\vec k - \vec q)\cdot \vec x}e^{-i(\vec l + \vec q)\cdot \vec y}e^{-i\vec q \cdot \vec D}\nonumber, 
\end{eqnarray}
where we recognize the inverse Fourier transforms of the $\delta$-functions:

\begin{eqnarray}
V(\vec D) & = & \int\frac{d \vec k d\vec l d\vec q}{(2\pi)^3}\frac{\tilde{\rho}_1(\vec k)\tilde{\rho}_2(\vec l)}{q^2}\delta(\vec k - \vec q)\delta(\vec l + \vec q)e^{-i\vec q \cdot \vec D}\nonumber \\
& = & \int\frac{d\vec q}{(2\pi)^3}\frac{\tilde{\rho}_1(\vec q)\tilde{\rho}_2(-\vec q)}{q^2}e^{-i\vec q \cdot \vec D}\nonumber
\end{eqnarray}
As the form factors $F_1$ and $F_2$ of the distributions $\rho_1$ and $\rho_2$ are by definition their Fourier transforms, this proves that the Fourier transform $\tilde V$ of the potential $V$ is given by:

\begin{equation}
\tilde V(\vec q)=\frac{F_1(\vec q)F_2(-\vec q)}{q^2}
\end{equation}

\section{Calculation of $V_{e\dot{e}^{+}}$}\label{appendixB}

Using \eqref{eq:4} and the expressions for the form factors $F_e$ and $F_{\dot{e}^{+}}$ of the visible and dark electronic clouds, we have:

\begin{equation}
V_{e\dot{e}^{+}}(r)=-\frac{Z\epsilon \alpha}{i\pi r}\sum\limits_{k=1}^3 c_k\int_{-\infty}^{\infty}\frac{dq}{q}\frac{256}{\left(4+\dot{a}_0^2q^2\right)^2\left(4+a_k^2q^2\right)^2}e^{iqr}\label{eq:28}
\end{equation}

This can be calculated by analytic continuation of the integrand in the complex plane and by residues. The function $f_k(q)=\frac{e^{iqr}}{q}\frac{256}{\left(4+\dot{a}_0^2q^2\right)^2\left(4+a_k^2q^2\right)^2}$ has poles in $q=0$ (order $1$), $q=\pm \frac{2i}{\dot{a}_0}$ (order $2$) and $q=\pm \frac{2i}{a_k}$ (order $2$) and can be rewritten as:

\begin{equation}
f_k=\frac{e^{iqr}}{q}\frac{256}{\dot{a}_{0}^{4}a_{k}^{4}\left(q-i\frac{2}{\dot{a}_0}\right)^2\left(q+i\frac{2}{\dot{a}_0}\right)^2\left(q-i\frac{2}{a_k}\right)^2\left(q+i\frac{2}{a_k}\right)^2}
\end{equation}

\begin{figure}
\begin{center}
\includegraphics[scale=0.25]{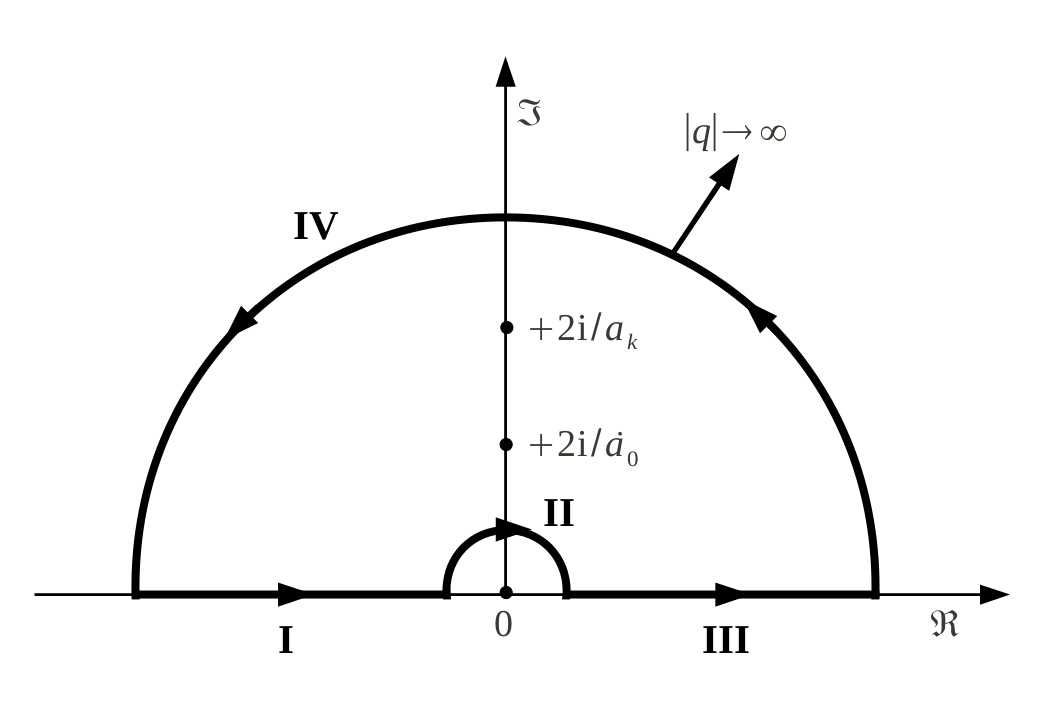}
\end{center}
\caption{Contour in the complex plane that encloses the poles of order $2$ in $q=+\frac{2i}{\dot{a}_0}$ and $q=+\frac{2i}{a_k}$ of the function $f_k(q)$. The limit of this contour for $|q| \rightarrow \infty$ on part IV is taken to calculate the integral \eqref{eq:28}.}
\label{contour}
\end{figure}

Taking the limit $|q| \rightarrow \infty$ of the contour represented in Figure \ref{contour}, we have:

\begin{equation}
\int_{\mathrm{I}}f_k dq+\int_{\mathrm{II}}f_k dq+\int_{\mathrm{III}}f_k dq+\int_{\mathrm{IV}}f_k dq=2i\pi \left(\mathrm{Res}_{f_k}\left(q=+\frac{2i}{\dot{a}_0}\right)+\mathrm{Res}_{f_k}\left(q=+\frac{2i}{a_k}\right)\right)\label{eq:30},
\end{equation}
with $\int_{\mathrm{IV}}f_k dq \rightarrow 0$ as the part IV of the contour is sent to regions where $\Im(q)>0$. Here, $\mathrm{Res}_{f_k}(q=z)$ denotes the residue of $f_k$ in $q=z$. Part II is realized clockwise and on half a turn around the pole in $q=0$, and thus:

\begin{equation}
\int_{\mathrm{II}}f_k dq=\left(-\frac{1}{2}\right)2i\pi \mathrm{Res}_{f_k}\left(q=0\right)\label{eq:31}
\end{equation}

The residues in $q=0$, $+\frac{2i}{\dot{a}_0}$ and $+\frac{2i}{a_k}$ are respectively given by:

\begin{eqnarray}
\mathrm{Res}_{f_k}\left(q=0\right) & = & 1\label{eq:32}, \\
\mathrm{Res}_{f_k}\left(q=+\frac{2i}{\dot{a}_0}\right) & = & e^{-2r/\dot{a}_0}\frac{T_k(r)}{4U_k}\label{eq:33}, \\
\mathrm{Res}_{f_k}\left(q=+\frac{2i}{a_k}\right) & = & e^{-2r/a_k}\frac{S_k(r)}{4U_k}\label{eq:34},
\end{eqnarray}
where $S_k(r)=(-2\dot{a}_0^4a_k^3+4\dot{a}_0^2a_k^5-2a_k^7)r-6\dot{a}_0^4a_k^4+8\dot{a}_0^2a_k^6-2a_k^8$ and $T_k(r)=(-2\dot{a}_0^3a_k^4+4\dot{a}_0^5a_k^2-2\dot{a}_0^7)r-6\dot{a}_0^4a_k^4+8\dot{a}_0^6a_k^2-2\dot{a}_0^8$ are polynomials of order $1$ in $r$ and $U_k=\dot{a}_0^8-4\dot{a}_0^6a_k^2+6\dot{a}_0^4a_k^4-4\dot{a}_0^2a_k^6+a_k^8$. Therefore, using \eqref{eq:30} to \eqref{eq:34}, we get for the integral of $f_k$ along the real axis:

\begin{eqnarray}
\int_{-\infty}^{\infty}f_k dq & = & \int_{\mathrm{I}}f_k dq+\int_{\mathrm{III}}f_k dq\nonumber \\
 & = & -\int_{\mathrm{II}}f_k dq+2i\pi \left(\mathrm{Res}_{f_k}\left(q=+\frac{2i}{\dot{a}_0}\right)+\mathrm{Res}_{f_k}\left(q=+\frac{2i}{a_k}\right)\right)\nonumber \\
  & = & i\pi \left(1+\frac{1}{2}\frac{e^{-2r/a_k}S_k(r)+e^{-2r/\dot{a}_0}T_k(r)}{U_k}\right)\label{eq:37}
\end{eqnarray}
and finally, inserting \eqref{eq:37} into \eqref{eq:28}, we get the desired expression for $V_{e\dot{e}^{+}}$:

\begin{equation}
V_{e\dot{e}^{+}}(r)=-\frac{Z\epsilon \alpha}{r}\sum\limits_{k=1}^3 c_k\left(1+\frac{1}{2}\frac{e^{-2r/a_k}S_k(r)+e^{-2r/\dot{a}_0}T_k(r)}{U_k}\right)
\end{equation}

\bibliographystyle{hope}
\bibliography{references}

\end{document}